\documentclass[ a4paper,floats,aps,nofootinbib,superscriptaddress,11pt, tightenlines, preprintnumbers]{revtex4}
\usepackage{graphicx}
\usepackage{amsmath,amssymb,amsfonts}
\usepackage{slashed}

\newcommand{\nue}{\tilde{\chi}_{1}^{0}}
\newcommand{\nueii}{\tilde{\chi}_{2}^{0}}
\newcommand{\nueiii}{\tilde{\chi}_{3}^{0}}

\newcommand{\svgg}{\langle \sigma v \rangle_{\gamma \gamma}}
\newcommand{\svgZ}{\langle \sigma v \rangle_{\gamma Z}}
\newcommand{\svann}{\langle \sigma v\rangle_{\rm{ann}}}

\begin{document}
\title{
Neutralino dark matter and the Fermi gamma-ray lines 
}

\preprint{IPPP/12/88}
\preprint{DCPT/12/176}
\preprint{TTP12-044}

\author{Guillaume~Chalons}
\email[]{guillaume.chalons@kit.edu}
\affiliation{Institut f\"{u}r Theoretische Teilchenphysik, Karlsruhe Institute of Technology, Universit\"{a}t Karlsruhe Engesserstra\ss e 7, 76128 Karlsruhe, Germany}
\author{Matthew J.~Dolan}
\email[]{m.j.dolan@durham.ac.uk}
\author{Christopher McCabe}
\email[]{christopher.mccabe@durham.ac.uk}

\affiliation{Institute for Particle Physics Phenomenology, Durham University, South Road, Durham, DH1 3LE, United Kingdom}

\begin{abstract} 
Motivated by recent claims of lines in the Fermi gamma-ray spectrum, we critically examine means of enhancing neutralino annihilation into neutral gauge bosons. The signal can be boosted while remaining consistent with continuum photon constraints if a new singlet-like pseudoscalar is present. We consider singlet extensions of the MSSM, focusing on the NMSSM, where a `well-tempered' neutralino can explain the lines while remaining consistent with current constraints. We adopt a complementary numerical and analytic approach throughout in order to gain intuition for the underlying physics. The scenario requires a rich spectrum of light neutralinos and charginos leading to characteristic phenomenological signatures at the LHC whose properties we explore. Future direct detection prospects are excellent, with sizeable spin-dependent and spin-independent cross-sections.

\end{abstract} 

\date{\today}
\maketitle

\section{Introduction}

There is robust evidence on astrophysical and cosmological scales for the presence of particle dark matter (DM) in our universe \cite{Jungman:1995df, Bertone:2004pz}. A particularly well studied candidate is the neutralino and determining the properties of this  particle is the subject of an intense experimental effort at colliders, direct detection and indirect detection experiments. Indirect detection experiments search for particles created from the decay or annihilation of DM particles from regions where the density of DM is expected to be high, such as the centre of our galaxy. A particularly striking signal is a monochromatic $\gamma$-ray line, arising when the DM annihilates into a two-body final state containing a photon \cite{Bergstrom:1988fp}.

Recently, numerous studies of the publicly available Fermi-LAT~\cite{Atwood:2009ez} data found a sharp feature in the $\gamma$-ray spectrum at $\sim 130$~GeV coming from the vicinity of the galactic centre~\cite{Bringmann:2012vr, Weniger:2012tx, Tempel:2012ey, Su:2012ft}. Interpreting the feature as a monochromatic line arising from DM annihilation into two photons, Weniger~\cite{Weniger:2012tx} found that DM with mass \mbox{$129.8\pm 2.4 ^{+7}_{-13}$~GeV} and annihilation cross-section \mbox{$\langle \sigma v \rangle_{\gamma\gamma} = \left(1.27\pm 0.32^{+0.18}_{-0.28}\right) \times 10^{-27} \mbox{ cm}^3 \mbox{s}^{-1}$} fits the signal well (see also~\cite{Tempel:2012ey, Su:2012ft,Hektor:2012kc,Su:2012zg,Hooper:2012qc, Mirabal:2012za, Hektor:2012jc,Zechlin:2012by}).  Intriguingly, refs.~\cite{Rajaraman:2012db,Su:2012ft} note that two lines, one at $\sim130$~GeV and a weaker one at $\sim111$~GeV, provide a slightly better fit to the Fermi data. Such a pair of lines can be naturally explained by a DM particle of mass \mbox{$\sim130$~GeV} annihilating into $\gamma \gamma$ and $\gamma Z$ with a relative annihilation cross-section $\langle \sigma v \rangle_{\gamma Z}/\langle \sigma v \rangle_{\gamma\gamma}=0.66^{+0.71}_{-0.48}$~\cite{Bringmann:2012ez}. In this paper, we address under what conditions the neutralino can fit these observations and the implications for the neutralino sector.

Doubts about a DM origin of these features have been raised in~\cite{Profumo:2012tr, Boyarsky:2012ca,Aharonian:2012cs}, although as yet there is no compelling astrophysical process that can explain the Fermi features. 
Searches for systematic effects associated with the Fermi-LAT instrument have been performed in \cite{Whiteson:2012hr, Hektor:2012ev, Finkbeiner:2012ez} and a small excess of photons from the Earth's limb at $\sim130$~GeV for photons within a limited range of detector incidence angles was found. However, such an effect cannot account for all of the signal from the galactic centre~\cite{Hektor:2012ev, Finkbeiner:2012ez}. Finally, the annihilation cross-section required to explain the feature is in mild tension with the upper limit $\svgg\lesssim1.0\times10^{-27} \text{ cm}^3 \text{s}^{-1}$ set by the Fermi-LAT Collaboration from the 2 year dataset~\cite{Ackermann:2012qk}. Furthermore, preliminary analysis of the 4 year dataset finds a feature at $\sim 135$ GeV but with a lower significance compared with the analyses quoted above~\cite{FERMI4}. The fate of the feature will ultimately be resolved as more data is collected. For the purposes of this paper, we assume that the features observed by Fermi are due to photons arising from neutralino annihilation and study the associated phenomenological implications.\footnote{Decaying DM is disfavoured as it requires an enhancement of the DM density near the galactic centre~\cite{Buchmuller:2012rc}.}

Interpreting the feature as a DM signature has its own challenges: ref.~\cite{Su:2012ft} find that the feature is offset from the galactic centre by $1.5^{\rm{o}}$, although this may arise due to an interplay between the galactic baryons and DM \cite{Kuhlen:2012qw}. A more serious issue is explaining the relatively large size of the annihilation cross-section into photons that is required while remaining consistent with the continuum flux of photons arising from annihilation of the DM into $W$ and $Z$ bosons and Standard Model fermions \cite{Buckley:2012ws, Buchmuller:2012rc, Cholis:2012fb,Cohen:2012me,Huang:2012yf}. As the DM is a neutral particle, its coupling to photons generally arises at loop-level and is suppressed relative to the coupling to $W$ and $Z$ bosons and Standard Model fermions, which can occur at tree-level. Cohen~et.~al.~\cite{Cohen:2012me} quantified this constraint through the quantity
\begin{equation}
R^{\rm{th}} \equiv \frac{\langle \sigma v\rangle_{\rm{ann}}}{2\langle \sigma v\rangle_{\gamma\gamma}+\langle \sigma v\rangle_{\gamma Z}}, 
\end{equation}
where $\langle \sigma v\rangle_{\rm{ann}}$ is the total annihilation cross-section and $\langle \sigma v\rangle_{\gamma\gamma,\gamma Z}$ are the annihilation cross-sections into $\gamma\gamma$ and $\gamma Z$ respectively. When the total annihilation cross-section is $\svann \sim 3 \times 10^{-26}$ $\rm{cm}^3\rm{s}^{-1}$ (required to achieve the correct thermal relic density) and when $\svgg \sim 1.2 \times 10^{-27}$ $\rm{cm}^3\rm{s}^{-1}$ and $\svgZ/\svgg=0.7$ (to explain the lines in the Fermi-LAT data), then $R^{\rm{th}} \sim 9$, compatible with all continuum constraints in~\cite{Cohen:2012me}. Thus, if the thermal relic density and line-strength for a given DM candidate are correct, it should be compatible with continuum photon constraints.\footnote{Similar conclusions can also be drawn from the constraints from $p\bar{p}$ and synchrotron radiation. Again, if the thermal relic density is correct, the DM candidate is compatible with these constraints, see e.g.~\cite{Buchmuller:2012rc,Belanger:2012ta} and~\cite{Laha:2012fg}.}

Many papers have been written with methods and models that explain the monochromatic features while remaining consistent with known constraints, see e.g.~\cite{Cline:2012nw,Lee:2012bq, Dudas:2012pb, Choi:2012ap, Kyae:2012vi,Buckley:2012ws,Chu:2012qy,Weiner:2012cb, Heo:2012dk, Frandsen:2012db,Park:2012xq,Tulin:2012uq,Cline:2012bz,Bai:2012qy,Bergstrom:2012bd, Weiner:2012gm, Fan:2012gr,Lee:2012wz,Wang:2012ts, D'Eramo:2012rr,Bernal:2012cd,Farzan:2012kk}. Within the framework of the simplest supersymmetric extension of the Standard Model, the MSSM, the large annihilation cross-section into photons $\svgg$ can be achieved~\cite{Boudjema:2005hb}. However, this requires a wino- or higgsino-like neutralino and such scenarios yield a negligible thermal relic density and also give rise to a large continuum flux~\cite{Acharya:2012dz}, which rule them out~\cite{Buchmuller:2012rc,Cholis:2012fb, Cohen:2012me}. Said another way, if the MSSM neutralino achieves the correct thermal relic density then $\svgg$ is orders of magnitude too low to explain the Fermi feature.\footnote{In fact, even in the MSSM the 130 GeV line can be explained by internal bremsstrahlung from a bino-like neutralino annihilating to light leptons~\cite{Bringmann:2012ez, Shakya:2012fj}. In order to enhance the number of photons produced, an accidental degeneracy between the mass of the exchanged t-channel slepton and the neutralino is required. We will not consider internal bremsstrahlung further in this paper.} Therefore, we see what is required: we need to keep the total annihilation cross-section at the level required to obtain the correct relic density while boosting the cross-section into photons. This can be achieved by exploiting a resonance with a pseudoscalar that couples primarily to photons, as suggested in~\cite{Ferrer:2006hy,Chalons:2011ia,Buckley:2012ws}. Although the MSSM does have a pseudoscalar Higgs $A$, this approach fails because $A$ does not couple primarily to photons; it couples to charged states $\bar{f} f$ at tree-level, while the coupling to photons is generated at loop level. However, as discussed in~\cite{Ferrer:2006hy,Das:2012ys,Kang:2012bq}, within extensions of the MSSM that include extra singlet states, the tree-level coupling to $\bar{f} f$ can be reduced while maintaining the coupling to photons.

Singlet extensions of the MSSM typically extend the Higgs and neutralino sectors of the MSSM and have received much attention as they can help to reduce the electroweak fine-tuning (see e.g.~\cite{Delgado:2010uj,Ross:2012nr, Ellwanger:2011mu, Hall:2011aa, Ross:2011xv,Asano:2012sv}). Indeed, refs.~\cite{Das:2012ys,Kang:2012bq} have shown that the correct annihilation cross-section into photons can be achieved in the ($\mathbb{Z}_3$-symmetric) NMSSM while achieving the correct thermal relic abundance and having a continuum photon flux consistent with observation. In those papers the neutralino is mostly bino, with a sub-dominant higgsino component. However, the benchmark point given in~\cite{Das:2012ys} predicts a neutralino-nucleon cross-section that is now excluded by the XENON100 Collaboration's search with 225 days of data~\cite{:2012nq}.

In this paper, we reexamine the ($\mathbb{Z}_3$-symmetric) NMSSM case in detail and again find that the neutralino is dominantly bino-higgsino-like.  As well as a numerical implementation, throughout, we strive for an analytic understanding of the phenomenology. In section~\ref{sec:nmssm} we discuss the general features that must be present for the neutralino to explain the lines and show that these are present within singlet extensions of the MSSM. We then consider the constraints from the requirement of obtaining the correct thermal relic density and from direct detection experiments. In particular, we describe how this scenario can be made consistent with the XENON100 bound. In section~\ref{sec:benchmark} we present three benchmark points with distinctive phenomenological signals and consider future tests of the model, including signals at the LHC. In section~\ref{sec:GNMSSM} we briefly discuss the neutralino phenomenology in a more general singlet extension of the MSSM that does not have a discrete $\mathbb{Z}_3$ symmetry before concluding in section~\ref{sec:conclusions}. Finally, details of the numerical procedure we follow to calculate $\svgg$ and $\svgZ$ as well as various analytic results are presented in appendices.

\section{Neutralino dark matter in singlet extensions of the MSSM}
\label{sec:nmssm}

The superpotential of the NMSSM, an extension of the MSSM by a gauge singlet chiral superfield $S$ and a discrete $\mathbb{Z}_3$ symmetry is\footnote{Our notation is such that $H_u\cdot H_d=H_u^+ H_d^- -H_u^0 H_d^0$.}
\begin{equation}
\mathcal{W}_{\rm{NMSSM}}=\mathcal{W}_{\rm{Yukawa}}+\lambda S H_{u}\cdot H_{d}+\frac{\kappa}{3}  S^3\;. 
\end{equation}
Here, $\mathcal{W}_{\rm{Yukawa}}$ is the usual MSSM superpotential containing the Yukawa couplings. The additional soft terms associated with the gauge singlet scalar are
\begin{equation}
\Delta V_{\rm{soft}} = m_S^2|S|^2+\left( \lambda A_{\lambda} S H_{u}\cdot H_{d} +\frac{\kappa A_{\kappa}}{3} S^3 +\rm{h.c.}\right)\;.
\end{equation}

In extensions of the MSSM by one singlet superfield, the neutralino mass matrix is diagonalised by an orthogonal real matrix $N$ in a similar fashion to the MSSM neutralino mass matrix. The resulting eigenvalues are then real, but not necessarily positive~\cite{Dreiner:2008tw}. The composition is
\begin{equation}
\nue \equiv N_{11} \tilde{B} + N_{12} \tilde{W} + N_{13} \tilde{H}_{d} + N_{14} \tilde{H}_{u}+N_{15} \tilde{S}\;.
\end{equation}
Following standard convention, we define the bino-fraction as $N_{11}^2$, the wino-fraction as $N_{12}^2$, the higgsino-fraction as $N_{13}^2+N_{14}^2$ and the singlino-fraction as $N_{15}^2$.  

\begin{figure}[t]
\centering
\begin{minipage}{0.5\textwidth}
\includegraphics[width=\textwidth]{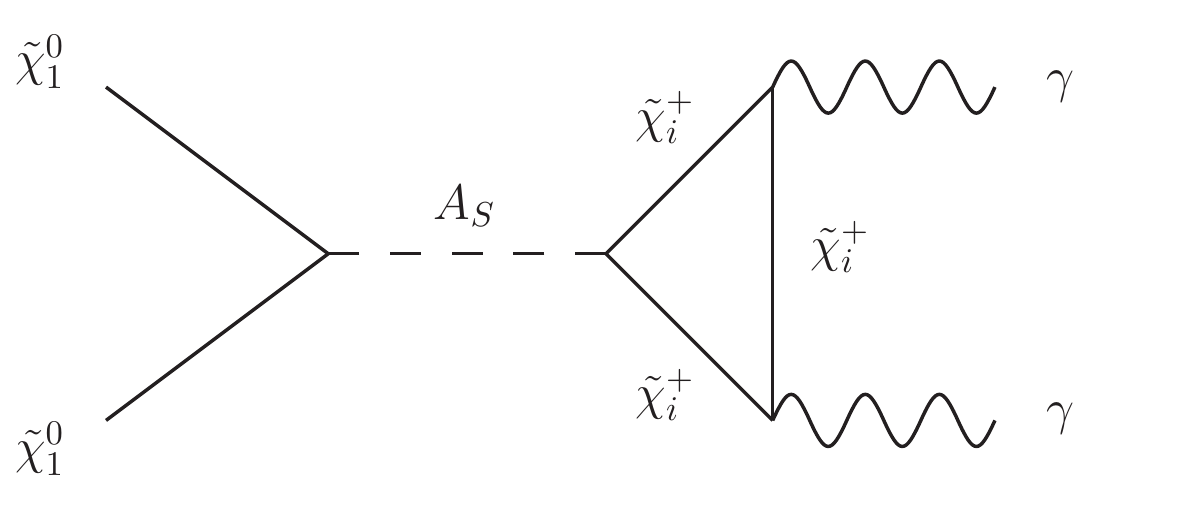}
\end{minipage}%
\begin{minipage}{0.5\textwidth}
\includegraphics[width=\textwidth]{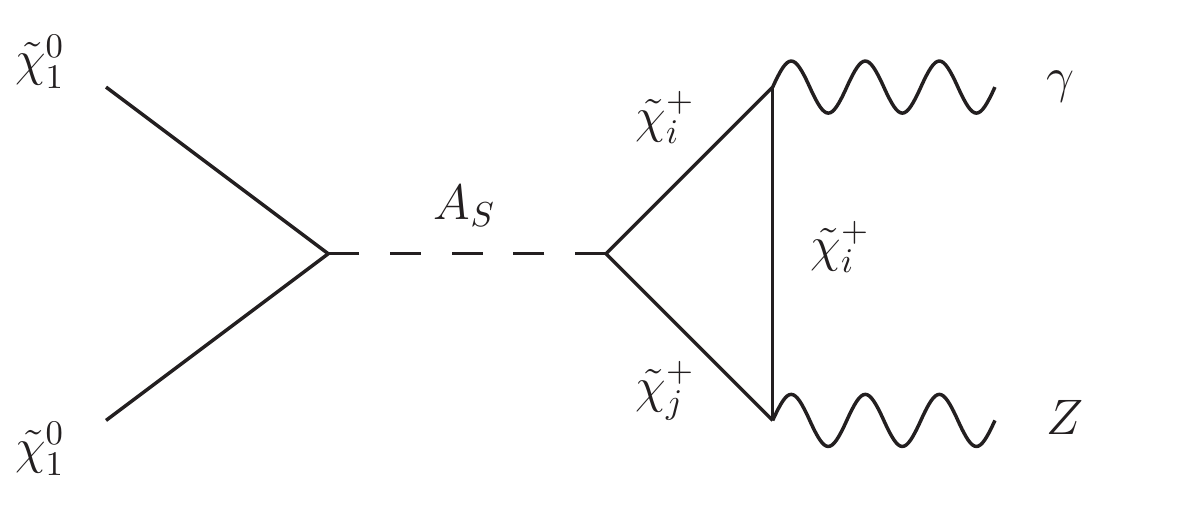}
\end{minipage}
\caption{The left and right panels show the dominant diagrams contributing to
$\svgg$ and $\svgZ$ when $2m_{\nue}\approx m_{A_S}$. The s-channel pseudoscalar Higgs $A_S$ is singlet-like in order to suppress the contribution to the continuum flux of photons. Both
charginos contribute to the process and in the right panel, the $Z$ coupling can mix contributions from different charginos $\chi_i^{\pm}$ and $\chi_j^{\pm}$.
\label{fig:schannel}}
\end{figure}

Boosting the $\gamma\gamma$ rate can now be easily achieved:  a pure-singlet pseudoscalar Higgs $A_S$ has no tree-level couplings to Standard Model fermions or gauge bosons. The dominant interaction is with neutralinos and charginos through the additional $\lambda S H_u\cdot H_d$ and $\kappa S^3/3$ superpotential terms. The charginos are necessarily heavier than the lightest neutralino so that, when $2 m_{\nue}\approx m_A$, the (non-relativistic) annihilation of neutralinos into photons is resonantly enhanced, through the diagrams shown in figure~\ref{fig:schannel}.

In the NMSSM, requiring a pseudoscalar with $m_{A_S}\approx2 m_{\nue}$ means that the singlet component of a neutralino with \mbox{$m_{\nue}\sim130$~GeV} is small: a singlet-like pseudoscalar Higgs $A_S$ is obtained in the NMSSM by requiring that the $(1,2)$ element of the CP-odd mixing matrix vanishes. This occurs if $A_{\lambda}\sim 2 \kappa s$. The resulting mass of the singlet-like pseudoscalar is (see e.g.~\cite{Ellwanger:2009dp} for further details)
\begin{equation}
m_{A_S}^2\sim -3 \kappa A_{\kappa} s=-3\, \frac{\kappa}{\lambda}\, A_{\kappa}\, \mu_{\rm{eff}}\;,
\end{equation}
where we have used $\mu_{\rm{eff}}=\lambda s$. If $A_{\kappa}\sim\mu_{\rm{eff}}\sim \mathcal{O}(100\text{ GeV})$, we require $\kappa/\lambda\sim1$ in order that $m_{A_S}\approx 2 m_{\nue}$. The singlino mass is
\begin{equation}
m_{\tilde{S}}=2 \kappa s=2 \left(\frac{ \kappa}{\lambda}\right) \mu_{\rm{eff}} \gtrsim  \mu_{\rm{eff}}
\label{eq:msinglino}
\end{equation}
and since this is larger than $\mu_{\rm{eff}}$, a neutralino with mass $\sim 130$ GeV will have a small singlino component (in the NMSSM examples we present below it is less than 1\%).\footnote{See~\cite{Das:2012ys} for a slightly different argument for why the singlino component should be small.}

\subsection{The monochromatic line from $\gamma\gamma$}

\begin{figure}[t]
\centering
\begin{minipage}{0.6\textwidth}
\includegraphics[width=\textwidth]{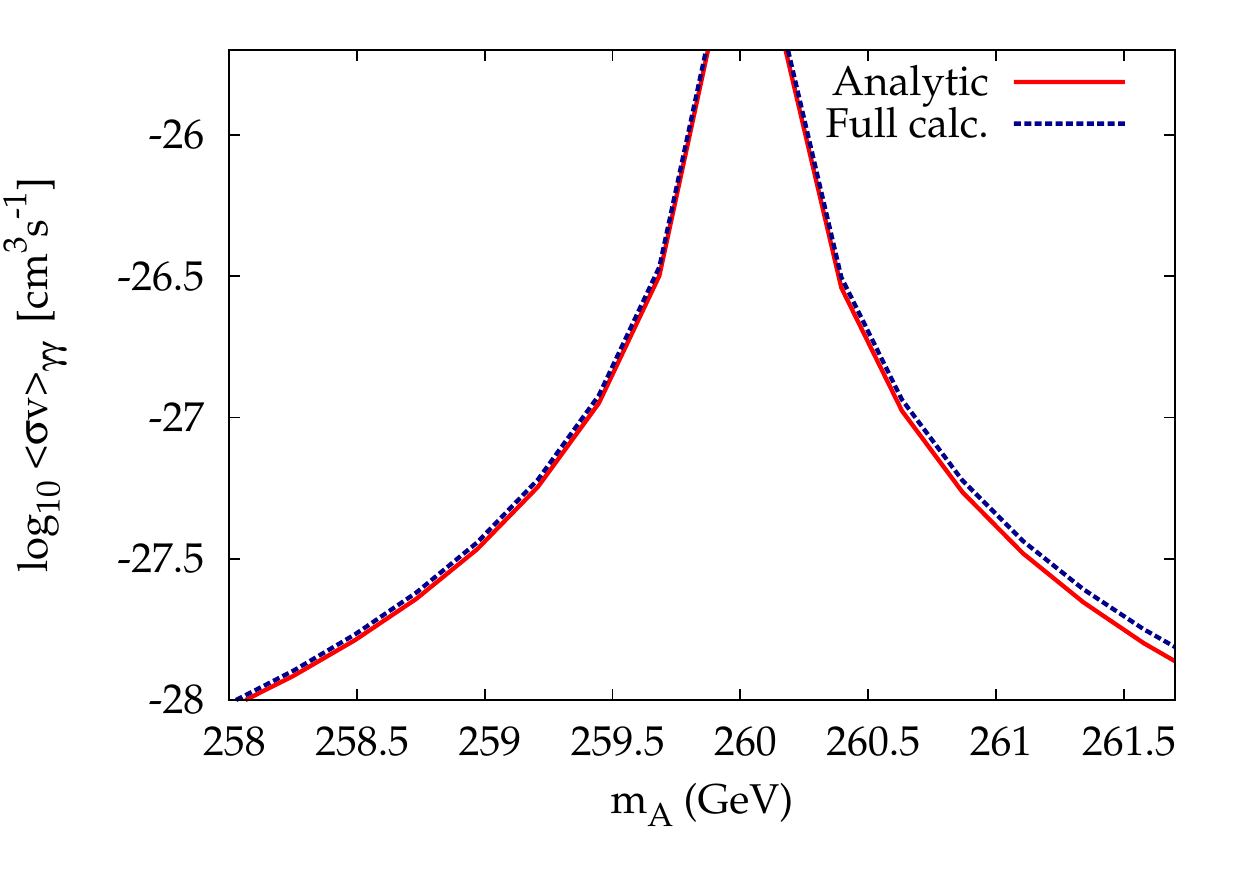}
\end{minipage}%
\caption{The annihilation cross-section $\svgg$ as a function of the pseudoscalar Higgs mass $m_A$ in the vicinity of our `well-tempered' benchmark point (discussed in more detail later). The solid red line is the result of our analytic formula (eq.~\eqref{eq:fullgg}) including only the diagrams with the pseudoscalar $A_S$. The dotted blue line shows the result of the full numerical calculation, detailed in appendix~\ref{sec:nmssm_calc}; the agreement is good. The width of the resonance at $\svgg\approx10^{-27}\text{ cm}^3{\rm{s}}^{-1}$ is $\sim 1$~GeV.
\label{fig:sigvcomp}}
\end{figure}

The annihilation cross-sections for $\chi\chi \to \gamma \gamma$ and $\chi\chi \to \gamma Z$ were computed in the MSSM in~\cite{Bergstrom:1997fh,Bern:1997ng} and adapted to the NMSSM in~\cite{Ferrer:2006hy}. We have implemented a complete numerical calculation of these cross-sections with the SloopS code~\cite{Boudjema:2005hb,Baro:2007em,Baro:2008bg,Baro:2009gn}, which we describe in appendix~\ref{sec:nmssm_calc}. In this section we discuss the process $\chi\chi \to \gamma \gamma$, leaving our discussion of $\chi\chi \to \gamma Z$ to section~\ref{Sec:gammaZ}. In order to gain intuition for these processes we also take an analytic approach. We have calculated the contribution to the cross-sections from the diagrams shown in figure~\ref{fig:schannel}, which dominate when $A_S$ is mostly singlet and near the resonance $m_A \approx 2 m_{\nue}$. The full expressions are given in appendix~\ref{sec:svan}. We compare the numerical and analytic results in figure~\ref{fig:sigvcomp} and find good agreement (this is for the `well-tempered' benchmark point, discussed in detail in section~\ref{sec:benchmark}). To simplify the full expression and highlight the parameter dependence of $\svgg$, we consider the result assuming that only the lightest chargino contributes
\begin{align}
\svgg &= \frac{\alpha^{2} \lambda^2  }{4\pi^3}\frac{\left( \lambda N_{13} N_{14} -  \kappa N_{15}^2\right)^2 \left(m_{\tilde{\chi}_1^+} U_{12} V_{12}\right)^2}{(4m_{\tilde{\chi}_1^0}^2-m_{A}^2)^2 + \Gamma_A^2 m_A^2} 
 \arctan^4\left( \sqrt{\frac{m_{\tilde{\chi}_1^0}^2}{m_{\tilde{\chi}_1^+}^2-m_{\tilde{\chi}_1^0}^2}} \right) \\
 &\sim 1.2\times 10^{-27}\text{ cm}^3{\rm{s}}^{-1}\; \left(\frac{\lambda}{0.7}\right)^2\left(\frac{\lambda N_{13} N_{14} -  \kappa N_{15}^2}{0.05}\right)^2\left(\frac{1.5\text{ GeV}}{\delta_{\chi}}\right)^2\;.
\label{eq:chigamgam}
\end{align} 
Here, $\alpha$ is the EM fine structure constant, $U$ and $V$ are the usual matrices that diagonalise the chargino mass matrix and $m_{\tilde{\chi}^+_i}$ is the chargino mass. In the second line, to further simplify the expression, we have assumed that the lightest chargino is degenerate in mass with the neutralino, $U_{12}=V_{12}=1$, $\Gamma_{A}=0$ and we have defined \mbox{$\delta_{\chi}=m_{A}-2 m_{\nue}$}. The numerical prefactor is the line-strength found in~\cite{Weniger:2012tx}.

In order to boost the signal without tuning the neutralino and pseudoscalar masses too much, we desire at least some of the following to be true:
\begin{itemize}
\item A large value of $\lambda$.
\item A large value of $\kappa$.
\item A neutralino with a large higgsino and/or singlino-fraction.
\item A light higgsino-like chargino with mass $m_{\tilde{\chi}_1^{\pm}} \approx m_{\nue}$.
\end{itemize}

In the NMSSM it is usual to keep $\lambda \lesssim 0.7$ in order that it remains perturbative up to the GUT scale. However, the singlet contribution to the lightest Higgs mass is proportional to $\lambda^2 \sin^2 2\beta$ so a larger value of $\lambda$ may reduce the electroweak fine tuning. This scenario in which part of the theory becomes non-perturbative has been dubbed $\lambda$-SUSY~\cite{Barbieri:2006dq,Hall:2011aa} and counter to naive expectation, it has been shown that gauge coupling unification may be improved~\cite{Hardy:2012ef}. We consider points with $\lambda \sim 0.7$ and above.
 
The pseudoscalar $A_S$ couples to the charginos through the $\lambda A_{S} \tilde{H}_u^+ \tilde{H}_d^-$ term so the signal is enhanced for a light higgsino-like chargino. Having a large higgsino-fraction for the neutralino helps in this respect as it implies that there should also be a higgsino-like chargino reasonably close in mass (in the limit of a pure higgsino-like neutralino, the neutralino and chargino are mass degenerate). Finally, we argued above that the singlino component of the neutralino is small in the NMSSM. Therefore, in this case, the higgsino component should be maximised while remaining consistent with constraints from the relic density and direct detection experiments. We address these points in turn.

\subsection{Thermal relic density}
\label{sec:relic}

\begin{figure}[t]
\centering
\begin{minipage}{0.6\textwidth}
\includegraphics[width=\textwidth]{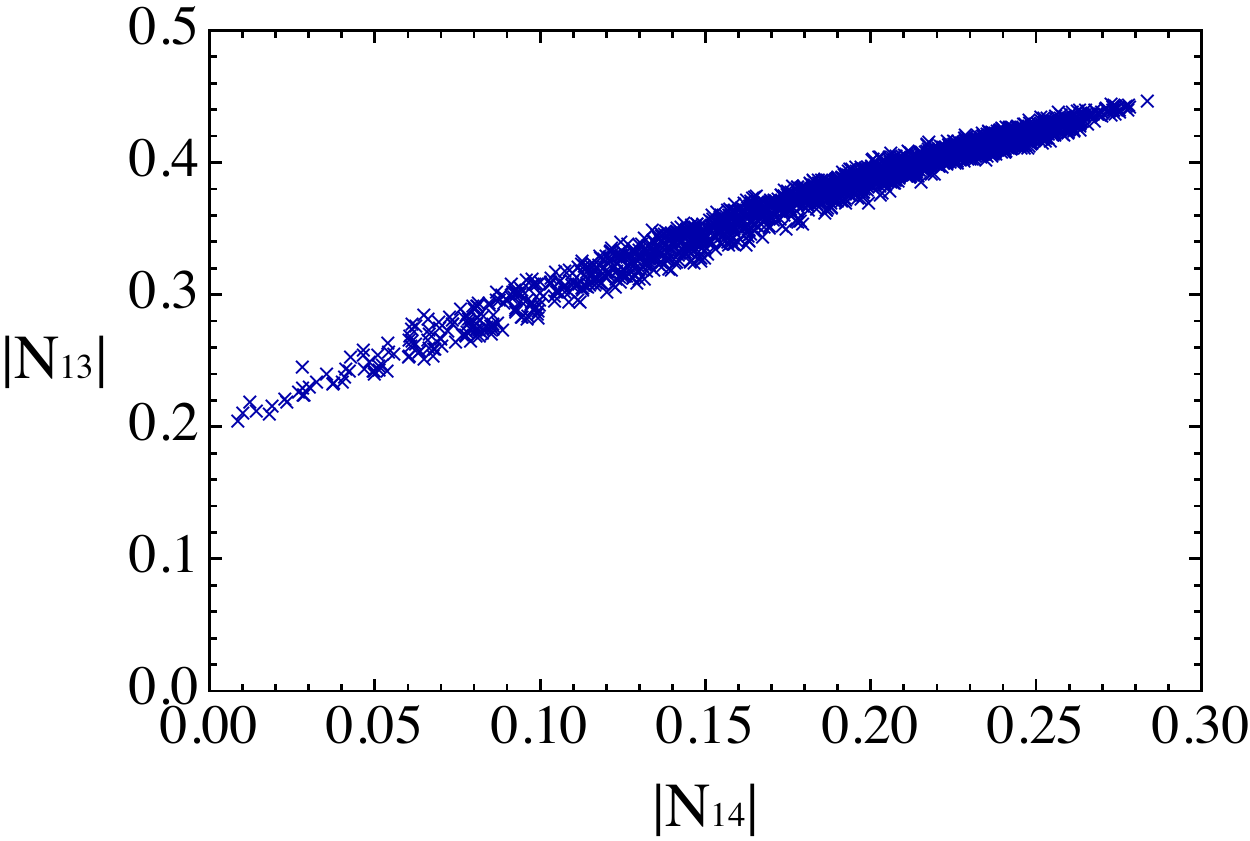}
\end{minipage}%
\caption{The results of a scan showing the values of $|N_{13}|$ and $|N_{14}|$ that give a thermal relic density in the range $0.09<\Omega_{\rm{DM}} h^2<0.13$ for a bino-higgsino like neutralino with $125<m_{\nue}<135$ GeV. The parameters that were scanned over are given in the text. The higgsino-fraction can be large, ranging from $\sim4-25\%$.
\label{fig:BHdensity}}
\end{figure}

In order for the lightest neutralino to be the DM, it should have the correct relic density. We assume that the neutralino makes up all of the DM and that it achieves it abundance through the usual mechanism of thermal freeze-out in the early universe~\cite{Zeldovich1,Zeldovich2,Chiu1,Lee:1977ua}. We have argued for a large higgsino-fraction so we first consider the thermal relic density of pure higgsino neutralino DM. To a good approximation this is given by \cite{ArkaniHamed:2006mb}
\begin{equation}
\Omega_{\rm{DM}}h^2\simeq0.1\left(\frac{\mu_{\rm{eff}}}{1 \text{ TeV}} \right)^2\;,
\end{equation}
where $m_{\nue}=\mu_{\rm{eff}}$ in this case. For a 130 GeV neutralino, the thermal relic density is too low for a pure higgsino to saturate the observed value. This is because higgsino neutralinos couple too efficiently to gauge bosons so in order to boost the relic density, we need to reduce the efficiency of this interaction. This can be achieved most easily by introducing a bino or singlino component. The wino component also couples efficiently to gauge bosons so this should be small in order to maximise the higgsino part. Therefore, to maximally enhance the $\svgg$ in the NMSSM, where the singlino component is small, the neutralino should be bino-higgsino-like. To gain intuition for how large the higgsino-fraction can be, in figure~\ref{fig:BHdensity} we show the values of $|N_{13}|$ and $|N_{14}|$ for which the observed relic density satisfies $0.09<\Omega_{\rm{DM}} h^2<0.13$ and $125<m_{\nue}<135$~GeV in the NMSSM. We have scanned over the parameters $-0.7< \lambda < 0.5$, $-1.1<\kappa<-0.3$, $1.8<\tan\beta<5$, $-170<\mu_{\rm{eff}} < -140$~GeV, $-100<A_{\kappa}<200$~GeV and $125<M_1<145$~GeV using \texttt{NMSSMtools~3.2.0}~\cite{Ellwanger:2004xm,Ellwanger:2005dv}, which uses \texttt{MicrOMEGAs}~\cite{Belanger:2005kh, Belanger:2006is,Belanger:2008sj,Belanger:2010pz} to calculate the relic abundance.\footnote{We have not included the $\chi\chi\to A_{S}\to \gamma\gamma$ process in our relic density calculation but it yields a subdominant contribution to the total annihilation cross-section that sets the relic abundance: $\svgg\approx0.03\langle \sigma v\rangle_{\rm{ann}}$.} We set $A_{\lambda}=2 \kappa s$ so that the pseudoscalar $A_S$ is singlet-like. This range captures points in the vicinity of our `well-tempered' benchmark point (see table~\ref{tab:benchmarks}). The higgsino-fraction can be large, ranging from $\sim 4-25\%$.

\subsection{Direct detection}


DM direct detection experiments search for low energy collisions between DM and a target nucleus. The low energy effective Lagrangian describing the neutralino-quark interaction is
\begin{equation}
\mathcal{L}_{\rm{eff}}= a_q \bar{\tilde{\chi}}_1^0 \nue \bar{q}q+d_q \bar{\tilde{\chi}}_1^0\gamma^{\mu}\gamma_5 \nue \bar{q}\gamma_{\mu}\gamma_5 q\;,
\end{equation}
where we only consider contributions that are not velocity or momentum-transfer suppressed. The first-term gives rises to a spin-independent interaction while the second gives rise to a spin-dependent interaction. The spin-independent interaction is mediated by CP-even Higgs exchange.\footnote{Throughout our discussion we work in the heavy squark limit and ignore the contribution from squark exchange.} In singlet extensions of the MSSM, the extra CP-even Higgs boson(s) and the $\lambda S H_u\cdot H_d$ and $\kappa S^3 /3$ superpotential terms lead to differences from the usual MSSM result for spin-independent scattering, as has been discussed in~\cite{Flores:1991rx,Bednyakov:1998is,Cerdeno:2004xw, Belanger:2005kh}. The extra CP-odd Higgs boson induces a new spin-dependent interaction, however, it is momentum-transfer suppressed~\cite{Freytsis:2010ne} and negligible compared to the dominant interaction mediated by the $Z$. 

The Lagrangian term $a_q$, responsible for the spin-independent interaction, is proportional to the neutralino-neutralino-Higgs coupling $g_{h_i \chi \chi}$, which in the NMSSM is given by
\begin{align}
\begin{split}
a_q\propto g_{h_i \chi \chi}&=g(N_{12}-\tan\theta_W N_{11})(S_{i1}N_{13}-S_{i2}N_{14})\\
&-\sqrt{2}\lambda(S_{i1}N_{14}N_{15}+S_{i2}N_{13}N_{15}+S_{i3}N_{13}N_{14})\label{Eq:ghchichilambda}\\
&+\sqrt{2}\kappa S_{i3} N_{15}^2\;.
\end{split}
\end{align}
The first line reduces to the MSSM result while the second and third lines come from the $\lambda S H_u \cdot H_d$ and $\kappa S^3 /3$ superpotential terms respectively. The index $i$ runs from one to three over the CP-even Higgs bosons, and $S$ is the matrix that diagonalises the Higgs mass-squared matrix. In appendix~\ref{sec:ddcalc}, we provide full details of how the scattering cross-section $\sigma_{\rm{SI}}$ is calculated from $a_q$.

For the case for a mostly bino-higgsino-like neutralino the MSSM-like contribution and $\lambda S H_u \cdot H_d$ contribution are dominant. We find that there is the possibility of a cancellation in the MSSM-like terms between the MSSM-like Higgs contributions when ${\rm{sgn}}(N_{13})={\rm{sgn}}(N_{14})$~\cite{Ellis:2000ds, Cohen:2010gj,Grothaus:2012js}.  Taking an explicit example, the `well-tempered' benchmark point in table~\ref{tab:benchmarks} has ${\rm{sgn}}(N_{13})={\rm{sgn}}(N_{14})$ and we find $\sigma_{\rm{SI}}^p=1.4\times10^{-9}$~pb. Repeating the calculation with the same parameters but with $N_{14}\rightarrow - N_{14}$, we find $\sigma_{\rm{SI}}^p=13.4\times10^{-9}$~pb, an order of magnitude larger. The current XENON100 limit is $\sigma_{\rm{SI}}^p=3.1\times 10^{-9} \text{ pb}$ for $m_{\nue}=130$ GeV~\cite{:2012nq} so we find that we require ${\rm{sgn}}(N_{13})={\rm{sgn}}(N_{14})$ to be phenomenologically viable.\footnote{The XENON100 limit is fairly robust against astrophysical uncertainties for DM in this mass range. See e.g.~\cite{McCabe:2010zh, Fairbairn:2012zs}.}  The benchmark point considered in~\cite{Das:2012ys} had $\rm{sgn}(N_{13})\neq \rm{sgn}(N_{14})$ and is now excluded by the 225 live-day XENON100 result. 

To understand when ${\rm{sgn}}(N_{13})={\rm{sgn}}(N_{14})$, we consider the MSSM tree-level result
\begin{equation}
\frac{N_{13}}{N_{14}}\approx \frac{-\mu_{\rm{eff}} (M_1-m_{\nue})+m_Z^2 \sin\beta \cos\beta \sin^2\theta_W}{m_{\nue} (M_1-m_{\nue})+m_Z^2 \cos^2\beta \sin^2\theta_W}\;,
\end{equation}
valid in the limit that $M_2$ is large. This relation is also approximately true in the NMSSM for the bino-higgsino case (and when $|\lambda|,|\kappa|\lesssim1$). In order that ${\rm{sgn}}(N_{13})={\rm{sgn}}(N_{14})$, we find
\begin{equation}
\mu_{\rm{eff}}<\frac{m_Z^2 \sin\beta \cos\beta\sin^2\theta_W}{(M_1-m_{\nue})}=88\text{ GeV} \left(\frac{5 \text{ GeV}}{M_1-m_{\nue}}\right)\left(\frac{\sin\beta \cos\beta}{0.23}\right)\;.
\end{equation}
For a mixed bino-higgsino-like neutralino, $M_1\sim |\mu_{\rm{eff}}|$ and $M_1\gtrsim m_{\nue}$. Thus, we must have $\mu_{\rm{eff}}<0$, since $|\mu_{\rm{eff}}| \leq 100~\rm{GeV}$ is ruled out by searches from LEP~\cite{Beringer:1900zz}. 

We next consider the spin-dependent interaction, which in the MSSM and its singlet extensions is mediated by the $Z$ boson. The Lagrangian term $d_q \propto |N_{13}|^2-|N_{14}|^2$, so a spin-dependent detection directly probes the higgsino component. In appendix~\ref{sec:ddcalc}, we show that the cross-section to scatter off a proton is
\begin{equation}
\sigma_{p}^{\rm{SD}}\approx 4.0 \times10^{-4}\text{ pb}\left( \frac{|N_{13}|^2-|N_{14}|^2}{0.1}\right)^2\;.
\end{equation}
The cross-section to scatter off a neutron has the same scaling but the pre-factor is $3.1 \times10^{-4}\text{ pb}$. 

Currently, the best limits on $\sigma_{p}^{\rm{SD}}$ are from Super-Kamiokande \cite{Tanaka:2011uf} and the combined IceCube and AMANDA-II limits \cite{IceCube:2011aj}, which have a similar sensitivity at 130 GeV, $\sigma_p^{\rm{SD}}\lesssim4\times10^{-4}$ pb. This limit assumes that the capture rate has reached equilibrium and the neutralino annihilates solely into $W^+ W^-$ in the Sun. This is a reasonably good approximation when the higgsino component is large. The expected sensitivity at $\sim130$ GeV of the completed IceCube detector is \mbox{$\sigma_p^{\rm{SD}}\approx 3\times10^{-5}$ pb}~\cite{IceCube:2011aj}. Direct limits can be placed from terrestrial direct detection experiments but their sensitivity is lower: for instance, COUPP set a limit of $\sigma_{p}^{\rm{SD}}=6\times10^{-3}$ pb at 130~GeV~\cite{Behnke:2012ys}. The strongest published limit on $\sigma_{n}^{\rm{SD}}$ is from ZEPLIN-III,  who find $\sigma_n^{\rm{SD}}<1\times10^{-2}$~pb at 130~GeV~\cite{Akimov:2011tj}. 
XENON100 should improve considerably on this with their 225 days dataset~\cite{:2012nq}: a recent independent analysis~\cite{Garny:2012it} obtains a limit of $\sigma_n^{\rm{SD}}\sim 9\times10^{-4}$~pb for the same mass.

\section{Benchmark points and future signatures}
\label{sec:benchmark}

We now present details of three benchmark points that satisfy all constraints. All points have a large enough value of $\svgg$ to explain the Fermi line at $\sim 130$~GeV, through a resonance with the pseudoscalar Higgs $A_S$. The important parameters for achieving the correct $\gamma$-ray line position and strength, as well as the those required to have a singlet-like pseudoscalar Higgs are listed in table~\ref{tab:benchmarks}, with other more incidental ones detailed in the caption. All points have small values of $\tan\beta$ to enhance the $\lambda^2$ tree-level contribution to the (lightest CP-even) Higgs mass; for all points $125 \text{ GeV}<m_{h_1}<126 \text{ GeV}$. We have also chosen all the points to have large values of $|\lambda|\geq 0.7$ and based on the results of~\cite{Hardy:2012ef}, are unconcerned about $\lambda$ or $\kappa$ becoming non-perturbative between tens of TeV and the GUT scale.

\begin{table}
\centering
\begin{tabular}{cccc} \hline 
& \raisebox{-0.5ex}{Well-} & \raisebox{-0.5ex}{Intermediate-} \\ [-1ex]
\raisebox{1.5ex}{Parameter}  & \raisebox{0.0ex}{Tempered} \, & \,\raisebox{0.0ex}{Slepton} \,&\,\raisebox{1.5ex}{$\lambda$-SUSY} \   \\[0.5ex] \hline  \\[-1.em]
$\lambda$ & -0.7 & -0.7 & -1.5 \\ 
$\kappa$ & -0.863 & -0.77  &-2.19 \\ 
$\tan\beta$ & 4.0 & 4.0 & 5.45 \\ 
$A_{\lambda}$~[GeV] & -369.9 & -378.0  &  -478.3\\ 
$A_{\kappa}$~[GeV] & 75.5 & 74.95 & -55.9 \\ 
$\mu_{\rm{eff}}$~[GeV] & -150.0 & -190.0& -168.0 \\ 
$M_1$~[GeV] & 135.0 & 135.5 & 128.4 \\[0.5ex] \hline  \\[-1.em] 

$m_{\nue}$~[GeV] & 130.0 & 133.7 & 129.9 \\ 
$N_{11}, \,N_{15}$ & -0.89, 0.1 & 0.96, -0.06 &  0.975, -0.083 \\ 
$N_{13},\,N_{14}$ & 0.39, 0.19 & -0.26, -0.09 & -0.21, 0.012 \\ 
$m_A$~[GeV] & 259.45 & 267.27 & 259.33 \\ 
$P_{13}$ & 0.9999 & 0.9999 & 0.9999 \\ 
$\delta_{\chi}=|m_A-2m_{\nue}|$~[GeV] & 0.55 & 0.13 & 0.47 \\ 
$m_{\tilde{\chi}^+_1} ,\,m_{\tilde{\chi}^+_2}$~[GeV] & -155.8, 727 & 196.8, -727 & 172.8, -727 \\ 
$U_{12},\,V_{12}$ & 0.999, -0.988 \,& 0.999, -0.988\, & 0.999, -0.988  \\ 
$m_{\nueii}, \,m_{\nueiii}$~[GeV] & 149.9, 180.4 & 190.4, 214.7 & 141.1, 239.6  \\ 
$m_{h_1}, \,m_{h_2},\, m_{h_3}$~[GeV] & 125.7, 310.8, 610.7 & 125.7, 366.2, 714.1 & 125.7, 327.3, 826.5 \\ 
$m_{\tilde{e}_R,\tilde{\mu}_R}$~[GeV]  & 900 & 144.7 & 901 \\[0.5ex] \hline \\[-1.em]

$\svgg\times 10^{-27} \text{ cm}^3 \text{s}^{-1}$ & 1.2  & 1.1 &0.9 \\ 
$R^{\rm{th}}$& 5.0  & 0.5 & 0.4 \\
$\Omega_{\rm{DM}} h^2$ & 0.10 & 0.12  & 0.11 \\ 
$\sigma_{\rm{SI}}^p\times 10^{-9}$~pb  & 1.4 & 0.23  & 3.1 \\ 
$\sigma_{\rm{SD}}^p\times 10^{-4}$~pb  & 5.4 & 1.4 & 0.7 \\ 
$\sigma_{\rm{SD}}^n\times 10^{-4}$~pb  & 4.2 & 1.1 & 0.5 \\ 
$\svgZ/\svgg$ & 0.64  & 0.52 & 0.67 \\ 
$\Delta a_{\mu}\times 10^{10}$ & $-1.0\pm2.9$ & $0.8\pm2.8$ & $-1.4\pm2.8$  \\[0.5ex] \hline
\end{tabular}
\caption{Properties and parameters of the benchmark points discussed in the text. All parameters have been defined in the text except $P_{13}$, the singlet component of $A_S$~\cite{Ellwanger:2009dp}. For all points the Higgs mass satisfies $125 \text{ GeV}<m_{h_1}<126 \text{ GeV}$. We fix $M_2=700$ GeV and $M_3=1.3$ TeV. Other soft parameters (at SUSY scale of 1 TeV): $A_t=-2.2$ TeV, $A_b=-1.0$ TeV, $A_{\tau}=A_{\mu}=0$, $m_{\tilde{L}_{e,\mu}}=m_{\tilde{Q}_t,\tilde{U}_t}=1.0$ TeV, $m_{\tilde{L}_{\tau},\tilde{E}_{\tau}}=0.8$ TeV, $m_{\tilde{Q}_{u,c},\tilde{U}_{u,c},\tilde{D}_{d}}=1.6$ TeV, $m_{\tilde{D}_b}=1.2$~TeV. 
 \label{tab:benchmarks}}
\end{table} 

The first point is a `well-tempered' point and has the largest higgsino-fraction $(19\%)$ of any of the points we consider. The large higgsino-fraction results in a large spin-dependent cross-section, sitting on the limit of $\sigma^{\rm{SD}}_p$ from IceCube and just below the limit of $\sigma^{\rm{SD}}_n$ from XENON100. Thus, this point should easily be detectable as more data is collected by those experiments.
The mass difference between the LSP $\nue$ and NLSP $\tilde{\chi}_1^+$ is small, at $\sim15$~GeV. With no intermediate mass sleptons, this point is also characterised by three-body decays of $\tilde{\chi}_2^0$ and $\tilde{\chi}_1^{+}$. This makes LHC searches for this point difficult, as we discuss in more detail in section~\ref{sec:collider}.

The second point is similar to the first but has an `intermediate-slepton' between $\nue$ and $\tilde{\chi}_1^+$. It is most like the benchmark point considered in~\cite{Das:2012ys}. Co-annihilation processes further deplete the relic density so the higgsino-fraction is smaller than before, resulting in a more massive $\tilde{\chi}_1^+$ and a higher fine-tuning between $m_A$ and $2 m_{\nue}$. We again find that the spin-dependent cross-section is relatively large and within reach of IceCube and XENON1T's projected sensitivity~\cite{Garny:2012it}. Interestingly, the spin-independent cross-section is small relative to the spin-dependent cross-section.  The intermediate mass sleptons allow the two-body decays of $\tilde{\chi}_2^{0}$ and $\tilde{\chi}_1^{+}$ to open up, resulting in promising LHC signatures (discussed in more detail in section~\ref{sec:collider}). 

The third point is a $\lambda$-SUSY style point with very large couplings: $\lambda=1.5$ and \mbox{$\kappa=-2.18$}. We have numerically solved the 1-loop RGEs for the $\lambda, \kappa$ system (including the effect of the top quark), and find that a Landau pole occurs at around 70~TeV. Counter to naive expectations, the fine-tuning between $m_A$ and $2 m_{\nue}$ is not significantly reduced. This is because $\lambda(\lambda N_{13} N_{14}-\kappa N_{15}^2)$ (c.f.~eq.~\eqref{eq:chigamgam}) is comparable to the value in the `well-tempered' point. The large value of $\lambda$ does increase the spin-independent cross-section through the $\lambda S H_u \cdot H_d$ terms (c.f.~eq.~\eqref{Eq:ghchichilambda}) and is just consistent with the current XENON100 limit. The lower higgsino-fraction compared to the previous points results in a lower spin-dependent cross-section, although still just within reach of IceCube's sensitivity.

It is well known that the Standard Model prediction for $a_{\mu}$ is more than $3\sigma$ below the experimental value~\cite{Hagiwara:2011af}. All of our benchmark points give a small additional contribution to the Standard Model value of $\Delta a_{\mu}$ but not enough to significantly bring the theoretical value into better agreement with the experimental result. The three benchmarks points have  $\mu_{\rm{eff}} <0$ while $M_1$ and $M_2$ are positive. In this case, ref.~\cite{Grothaus:2012js} has shown that the SUSY contribution to $(g-2)_{\mu}$ can be increased if the bino-higgsino-$\tilde{\mu}_R$ contribution dominates and there is a large hierarchy between the left- and right-handed sleptons. An alternative possibility for achieving a consistent value is through the introduction of flavour violation~\cite{Moroi:1995yh} but we do not consider this further.

\subsection{A second line from $\gamma Z$} \label{Sec:gammaZ}


There is strong evidence for a $\gamma$-ray line at $\sim130$~GeV and intriguingly, there is also weaker evidence for a second line at $\sim111$ GeV~\cite{Rajaraman:2012db,Su:2012ft}. The best fit to the relative annihilation cross-section is \mbox{$\langle \sigma v \rangle_{\gamma Z}/\langle \sigma v \rangle_{\gamma\gamma}=0.66^{+0.71}_{-0.48}$}~\cite{Bringmann:2012ez}. As we have already noted, this model does predict a second line with energy 
\begin{equation}
E_{\gamma}=m_{\nue}\left(1-\frac{m_{Z}^2}{4 m_{\nue}^2} \right)\;
\end{equation}
through the process shown in the right panel of figure~\ref{fig:schannel}. For $m_{\nue}=130$~GeV, the second line has \mbox{$E_{\gamma}=114$}~GeV. In table~\ref{tab:benchmarks} we quote the values for $\svgZ/\svgg$ for our benchmark points, finding that they are all consistent with the measured value. The numbers for $\svgZ$ quoted in the table were calculated numerically in SloopS and include the field renormalisation $\delta Z_{\gamma Z}^{1/2}$, important when the higgsino-fraction is large~\cite{Boudjema:2005hb}. This contribution is missing in~\cite{Das:2012ys,SchmidtHoberg:2012ip}. Further technical details are given in appendix~\ref{sec:nmssm_calc}.

\begin{figure}[t]
\centering
\begin{minipage}{0.6\textwidth}
\includegraphics[width=\textwidth]{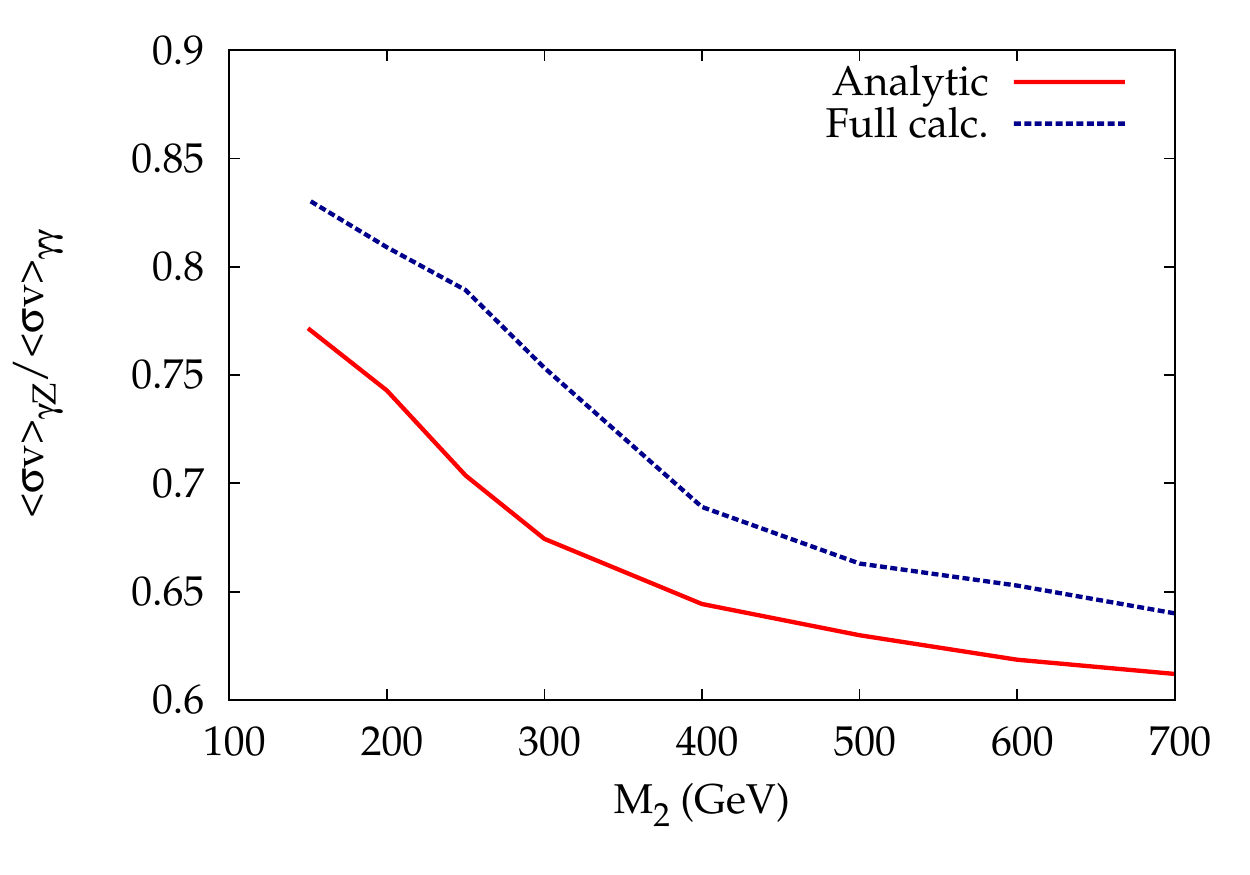}
\end{minipage}
\caption{The ratio of the $\gamma Z$ to $\gamma\gamma$ annihilation cross-sections near the `well-tempered' benchmark point as $M_2$ is varied. We have also varied $\mu_{\rm{eff}}$ to keep the relic density constant and $m_{\nue}=130$~GeV. The solid red line is the result of our analytic formula (eq.~\ref{eq:chigamz}) and the dotted blue line is the result of the full numerical calculation. The $\sim 10\%$ difference is due to the field renormalisation $\delta Z_{\gamma Z}^{1/2}$, missing from the analytic result. The ratio increases as $M_2$ is lowered, a result of the increasing wino component in the lighter chargino.
\label{fig:gZgg}}
\end{figure}

In appendix~\ref{sec:svan} we also present an analytic calculation of $\svgZ$, adapting the MSSM result from~\cite{Ullio:1997ke}. Taking the ratio of our analytic results for $\svgg$ and $\svgZ$, we find
\begin{equation}
\frac{\svgZ}{\svgg}=\frac{1}{8 \sin^2 2\theta_W}\left( 1-\frac{m_Z^2}{4 m_{\nue}^2}\right) \frac{\mathcal{F}^2_{\tilde{\chi}^{\pm}}}{\mathcal{G}^2_{\tilde{\chi}^{\pm}}}
\label{eq:chigamz}
\end{equation}
where $\mathcal{F}_{\tilde{\chi}^{\pm}}$ and $\mathcal{G}_{\tilde{\chi}^{\pm}}$ are functions that depend on the chargino masses and mixing matrices: $m_{\tilde{\chi}_1^+},m_{\tilde{\chi}_2^+},U,V$. The full expressions can be found in appendix~\ref{sec:svan}. From this formula, we see that once $m_{\nue}$ is fixed (in our case, through the position of the $\gamma\gamma$ line) and since $m_Z$ and $\sin\theta_W$ are known, the ratio of $\svgZ/\svgg$ directly probes properties of the chargino sector.

To demonstrate this, in figure~\ref{fig:gZgg} we take our `well-tempered' benchmark point and show how the ratio $\svgZ/\svgg$ changes as we vary the parameter $M_2$. As $M_2$ decreases, we increase $\mu_{\rm{eff}}$ in order that $m_{\nue}$ and $\Omega_{\rm{DM}}h^2$ remain fixed. We also adjust $A_{\lambda}$ and $A_{\kappa}$ so that $A_S$ remains singlet-like with mass $m_A\approx 2 m_{\nue}$. The solid (red) line is the result of our analytic formula eq.~\eqref{eq:chigamz} and the dotted (blue) line is the result of the full numerical calculation. The $\sim 10\%$ difference is due to the field renormalisation $\delta Z_{\gamma Z}^{1/2}$, missing from the analytic result.

Smaller values of $M_2$ lead to an increase in the wino-fraction of the lighter chargino and a smaller mass for the heavier chargino.  The chargino-chargino-$Z$ coupling is larger for a wino-like chargino compared with a higgsino-like chargino. Therefore, as $M_2$ is lowered, this coupling grows for the lighter chargino while the chargino-chargino-$\gamma$ coupling remains constant (since it is simply the electric charge) resulting in an increase in the ratio $\svgZ/\svgg$.

\subsection{LHC signatures}
\label{sec:collider}

Although the singlet-like pseudoscalar Higgs with $m_A\approx2 m_{\nue}$ is essential, it does not give rise to any observable collider signature. Therefore, the best prospects for observation of new physics at the LHC in this scenario are in searches sensitive to neutralino and chargino production. All of our benchmark points have a relatively large higgsino-fraction in the LSP $\nue$. This implies that there will be light $\tilde{\chi}_{2,3}^0$ and $\tilde{\chi}_1^{\pm}$, which are mostly higgsino (as shown in table~\ref{tab:benchmarks}). If there are no intermediate sleptons or squarks in the spectrum, these will decay via three-body decays. On the other hand, when there is a slepton whose mass lies between that of the LSP and the higgsinos, the higgsinos will decay to the slepton and then to the LSP via two-body decays. In both these cases the most promising signatures involve dileptons and trileptons plus missing energy carried away by the LSP.

\begin{figure}[t]
\begin{minipage}{0.5\textwidth}
\includegraphics[width=\textwidth]{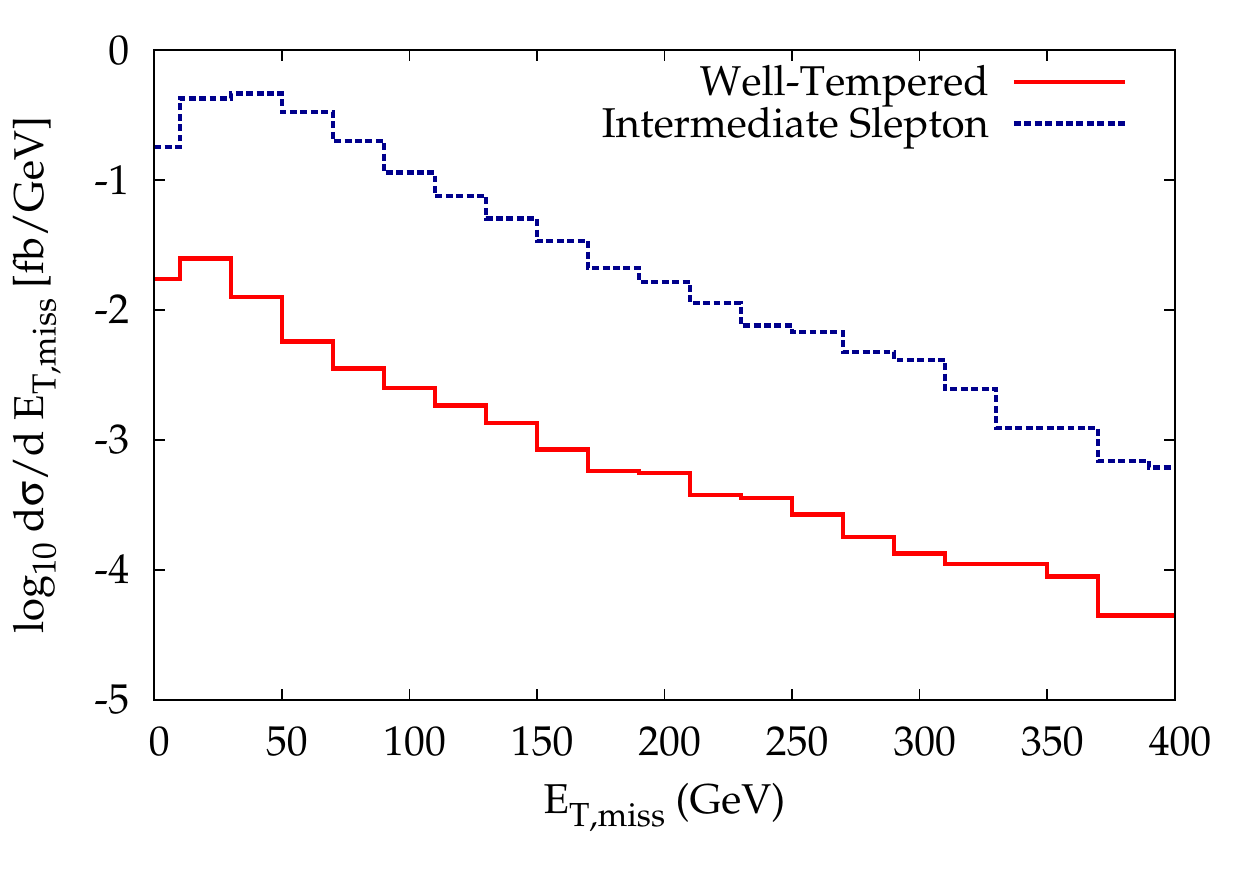}
\end{minipage}%
\begin{minipage}{0.5\textwidth}
\includegraphics[width=\textwidth]{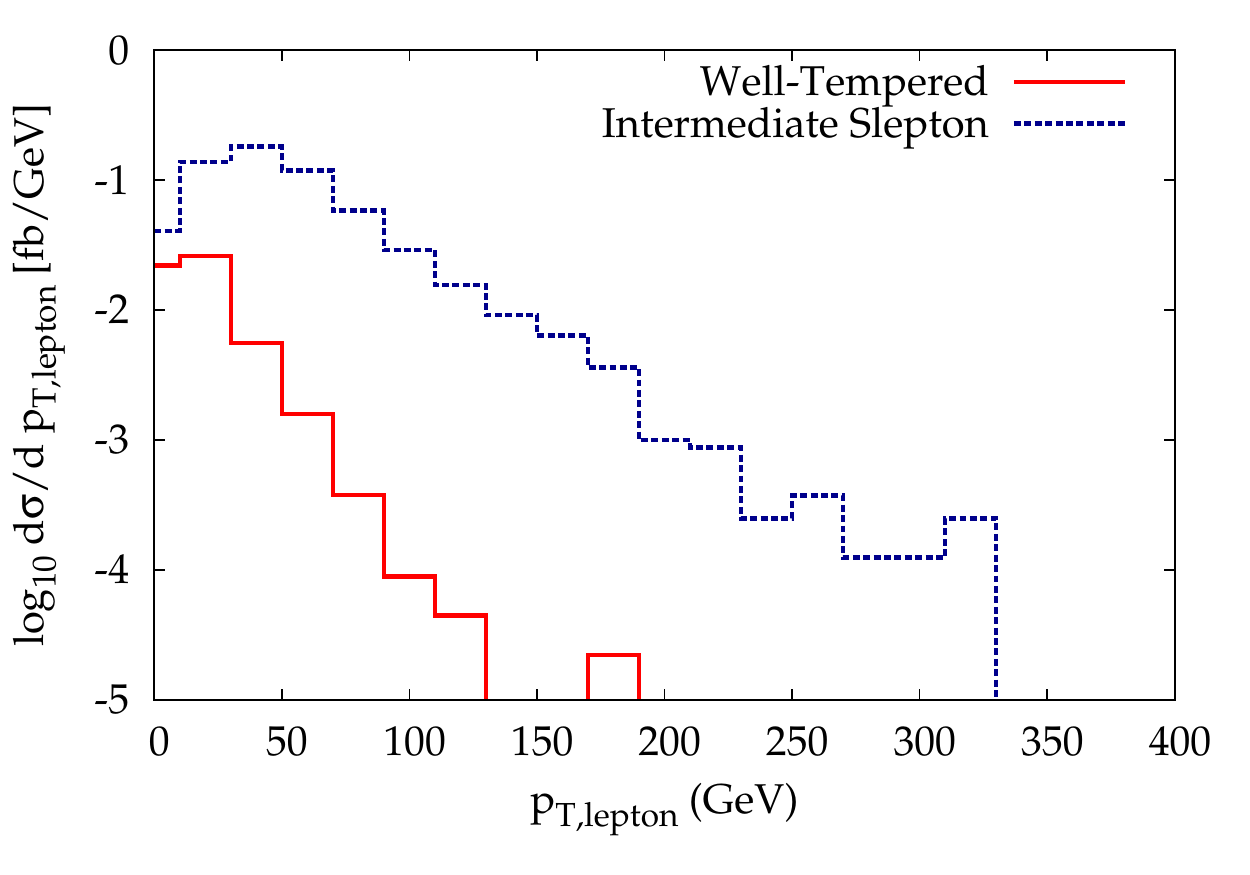}
\end{minipage}
\caption{Left panel: the missing energy distribution for our `well-tempered' and `intermediate-slepton' benchmark points for trilepton events at 14~TeV. The distributions have a similar shape as the neutralino mass is the same, but the `well-tempered' one has a smaller cross-section. Right panel: the $p_T$ distribution of the leading lepton (chosen to be an electron) for the same event selection. The more compressed `well-tempered' spectrum is much softer than the `intermediate-slepton' case. A targeted search with lower leptonic $p_T$ triggers may be helpful these cases. The distributions from the `$\lambda$-SUSY' benchmark point (not shown) are similar to the `well-tempered' distributions.
\label{fig:LHC}}
\end{figure}

Both ATLAS~\cite{:2012ku,:2012gg} and CMS~\cite{cmssearch} have recently published searches for direct gaugino production at $\sqrt{s}=7$~TeV with these final states.
We have implemented the ATLAS two- and three-lepton searches in the \texttt{RIVET} framework~\cite{Buckley:2010ar} (we expect the implications of the CMS results to be similar). We generate events using \texttt{Herwig++}~2.5.2~\cite{Bahr:2008pv,Gieseke:2011na} and implement the experimental event selection using \texttt{RIVET}~1.8.1 to analyse the fully showered final state. The ATLAS dilepton search~\cite{:2012gg} does not claim sensitivity to spectra where the gauginos decay via three-body decay without the presence of an intermediate slepton. Our results are in agreement with this. 
We find that the ATLAS search has a low sensitivity to our `well-tempered' and `$\lambda$-SUSY' benchmark points due to the small mass splitting ($\sim16$~GeV and $\sim11$~GeV respectively) between the LSP and the higgsino-like charginos $\tilde{\chi}_1^{\pm}$ and neutralinos $\tilde{\chi}_{2,3}^0$. This small splitting means that the transverse momentum $p_T$ of the leptons is generally low, leading to a small acceptance in all search channels.  Our benchmarks are also not ruled out by the trilepton searches. While we were finishing this article new gaugino searches from ATLAS and CMS based on data taken at 8~TeV were announced at the HCP2012~\cite{ATLAS-CONF-2012-154,CMSgauginos} conference. These results extend the reach with respect to $m_{\tilde{\chi}^{\pm}}$ but do not close the gap at low $m_{\tilde{\chi}^{\pm}}-m_{\tilde{\chi}^0}$, which may require data at 14~TeV or a targeted analysis.

To highlight the phenomenological differences between the `well-tempered' and `intermediate slepton' cases, in figure~\ref{fig:LHC} we show the missing energy distribution (left panel) and $p_T$ distribution of the the leading electron for the `well-tempered' point (in red, solid line), and the `intermediate-slepton' point (in dark blue, dashed line) at an energy of 14~TeV. The `$\lambda$-SUSY' distributions are similar to the `well-tempered' distributions as both originate from three-body decays of $\tilde{\chi}_{2,3}^0$ and $\tilde{\chi}_1^{\pm}$ (which occur because there is no intermediate mass slepton) and the small mass splitting between the LSP and NLSP. We therefore omit the `$\lambda$-SUSY' point from the plots. We have imposed the trigger cuts detailed in~\cite{:2012gg}, along with lepton isolation and associated quality criteria. Jets are reconstructed using the anti-$kt$ algorithm~\cite{Cacciari:2008gp}, with $R=0.4$, and we only accept points that yield exactly three leptons. As both points have a neutralino with the same mass, the shape of the MET distribution is similar. However the cross-section for the well-tempered case is lower due to the smaller probability that three lepton pass the $p_T$ cuts. In the right-hand panel of figure~\ref{fig:LHC} we show the $p_T$ spectrum for the leading lepton when it is an electron; the plot is similar for the muon case. The spectrum for the `intermediate-slepton' point peaks at higher $p_T$ and has a much longer tail than the `well-tempered' point, as there is more phase space for the leptons in this case. Probing the `well-tempered' region will prove difficult at the LHC, and may require a dedicated search with lower leptonic $p_T$ triggers than are currently implemented. One possibility for constraining this scenario might be via a multilepton search, where the pair-produced gaugino system recoils off an ISR-jet, leading to three collimated leptons recoiling against a jet. A second possibility is to look at a dilepton plus $\gamma$ search since the decay $\tilde{\chi}_2^0 \to \tilde{\chi}_1^0 \gamma$ can be large when $\tilde{\chi}^0_2$ decays otherwise only by three-body decay~\cite{Haber:1988px}. For instance, for the `well-tempered' benchmark point, this branching ratio is 47\%. We do not pursue these options further at the present time. It is also worth noting that the position of the $\gamma$-ray line in the Fermi data tells us the neutralino mass, so a very accurate reconstruction of the chargino and slepton masses may be possible using $M_{T2}$-based techniques~\cite{Lester:1999tx,Barr:2003rg}, for example.

Even though it is difficult to constrain the `well-tempered' scenario at the LHC, it is interesting to note that there is a complementarity between the LHC and direct detection searches for our benchmark points. 
The `well-tempered' benchmark point is difficult to constrain at the LHC as the high LSP higgsino-fraction leads to a compressed spectrum. However, the large higgsino component of the LSP in this case will be directly probed at the next generation of direct detection experiments. On the other hand,  when the higgsino component of the LSP is lower and the nucleon scattering cross-section is smaller, the splitting between the LSP and higgsinos is larger, leading to a harder $p_T$ spectrum. Also, the intermediate slepton required to deplete the relic density leads to more leptons being produced in decay chains. Thus, both these scenarios are amenable to testing in the near future.

\section{Extensions beyond the NMSSM}
\label{sec:GNMSSM}

The benchmarks points that we have discussed have all been within the NMSSM, which has a discrete $\mathbb{Z}_3$ symmetry. However, our analytic results did not rely on this symmetry so will apply more generally. Therefore, we briefly consider the phenomenology of a more general singlet extension of the MSSM. The most general extension of the MSSM by a gauge singlet superpotential is
\begin{equation}
\mathcal{W}= \mathcal{W}_{\rm{NMSSM}} + \mu H_{u}\cdot H_{d}+\xi S +\frac{1}{2}\mu_S S^2\;.
\end{equation}

One strong motivation for the NMSSM is that it solves the $\mu$-problem of the MSSM so including an explicit $\mu$-term in the superpotential naively ruins this solution. However, the superpotential with $\xi=0$ (the linear term in $S$ can be removed by a shift in its vev) and an underlying $\mathbb{Z}_4^R$ symmetry leads to $\mu\sim\mu_S \sim \mathcal{O}(m_{3/2}$) when the $\mathbb{Z}_4^R$ symmetry is broken to the usual matter parity after SUSY breaking.\footnote{This $R$-symmetry also cures the domain wall problem of the NMSSM~\cite{Abel:1995wk, Abel:1996cr, Panagiotakopoulos:1998yw}.}  This has been coined the GNMSSM~(see~\cite{Lee:2010gv, Lee:2011dya} for further details). We briefly consider the effect of these additional terms.

In section~\ref{sec:nmssm}, we argued that the singlino component of the neutralino in the NMSSM is small for \mbox{$m_{\nue}\sim130$~GeV}. In the GNMSSM, there is an extra contribution to $\mu_{\rm{eff}}$ and $m_{\tilde{S}}$ from $\mu$ and $\mu_S$ respectively. With these extra parameters, the singlino mass is
\begin{equation}
m_{\tilde{S}}=2 \kappa s+\mu_S=2  \left(\frac{ \kappa}{\lambda}\right) \mu_{\rm{eff}} +\left(\mu_S-2\frac{\kappa}{\lambda} \mu\right)\;.
\end{equation}
Comparing with eq.~\eqref{eq:msinglino}, we see that we now have the freedom to choose $\mu$ and $\mu_S$ such that $m_{\tilde{S}}\lesssim\mu_{\rm{eff}}$ and therefore, have a neutralino with \mbox{$m_{\nue}\sim130$~GeV}  where the singlino component is large.

For a singlino-higgsino neutralino, we can estimate the size of the singlino component required to obtain the correct relic density from figure~\ref{fig:BHdensity}. This is because, for heavy squarks and sleptons, the bino component acts like a singlino component relative to the much more efficient higgsino component. From figure~\ref{fig:BHdensity}, we take $N_{13}\sim0.3$, $N_{14}\sim0.1$ and $N_{15}\sim0.95$ in order to estimate the expected fine tuning required between $m_{\nue}$ and $m_A$. In eq.~\eqref{eq:chigamgam}, we considered $(\lambda N_{13} N_{14}-\kappa N_{15}^2)=0.05$ and found $\delta_{\chi}\sim1.5$ GeV. Using $\lambda\approx\kappa\approx0.7$ and the numbers above, we find that $(\lambda N_{13} N_{14}-\kappa N_{15}^2)\sim0.6$, larger by a factor of $\sim10$. As a result, we expect that $\delta_{\chi}$ is about an order of magnitude larger and the resulting tuning between $m_{\nue}$ and $m_A$ should be less severe in the GNMSSM. As this work was nearing completion, a dedicated study of the GNMSSM scenario confirms these findings~\cite{SchmidtHoberg:2012ip}.

\section{Conclusions}
\label{sec:conclusions}

Motivated by recent claims of lines in the Fermi gamma-ray spectrum, we have investigated enhancing neutralino annihilations into photons in singlet extensions of the
MSSM. The enhancement occurs when the lightest neutralino is on resonance with a singlet-like pseudoscalar
$A_S$ (see figure~\ref{fig:schannel}). Throughout, we have adopted an analytic approach as well as a numerical approach to understand the underlying physics. Using the analytic approach, it is clear that the lightest neutralino must have a large higgsino or singlino component in order to couple efficiently to this resonance through the $\lambda S H_u\cdot H_d$ or $\kappa S^3 /3$ superpotential terms.

We focused particularly on the phenomenology in the NMSSM, presenting benchmark points in table~\ref{tab:benchmarks} that are consistent with all constraints, including the continuum flux of photons and the relic density. In section~\ref{sec:nmssm}, we argued that the neutralino singlino-fraction is low in the NMSSM. The higgsino-fraction is bounded from above by the requirement of achieving the correct relic density but can still be as high as 25\% (see figure~\ref{fig:BHdensity}).  A large higgsino component is correlated with a large nuclear scattering cross-section; both the spin-independent and spin-dependent cross-sections are large. In order to ensure that the spin-independent cross-section is below the current limit set by XENON100, it is necessary to exploit cancellations between contributions from different CP-even Higgs bosons. This occurs when $\mu_{\rm{eff}} < 0$.

In this scenario a $\gamma Z$ line at energy $\sim 114$ GeV accompanies the $\gamma\gamma$ line. The relative strength of these lines depends on the chargino masses and mixings (see figure~\ref{fig:gZgg}). Currently the relative strength is poorly known but a more precise determination will help to constrain properties of the chargino sector. 

Due to the large higgsino-fraction of the lightest neutralino, in general there is a light higgsino-like chargino and two other higgsino-like neutralinos close in mass to the lightest neutralino. Decays of these particles give rise to LHC signatures with dileptons and trileptons plus missing energy. We investigated these signatures for two of our benchmark points; one with a slepton with a mass between the lightest chargino and lightest neutralino (`intermediate-slepton') and one without (`well-tempered') (see figure~\ref{fig:LHC}). With an intermediate slepton, the LHC signatures are promising. The case without an intermediate slepton is more difficult to probe at the LHC and may require a dedicated search with lower leptonic $p_T$ triggers than are currently implemented.

If a 130 GeV neutralino is the source of the gamma-ray features observed by Fermi, the associated phenomenology at colliders and direct detection experiments is potentially very rich. 

\section*{Acknowledgements}
\noindent MD and CM thank C\'eline B\oe hm, Felix Kahlh\"ofer and Kai Schmidt-Hoberg for discussions. GC would like to
thank the IKTP TU Dresden for the warm hospitality while parts of this work
were carried out.  The work of
GC is supported by BMBF grant 05H12VKF.


\appendix
\section{Numerical implementation of $\svgg$ and $\svgZ$}
\label{sec:nmssm_calc}

In the NMSSM, $\svgg$ and $\svgZ$ were computed
in~\cite{Ferrer:2006hy,Das:2012ys} by adapting the formulas of~\cite{Bergstrom:1997fh,Ullio:1997ke} to the NMSSM case. A complete numerical
calculation for these final states has also been performed in~\cite{Chalons:2011ia,Chalons:2012hf}. The calculation of the
$\gamma Z$ final state requires the calculation of the field
renormalisation $\delta Z^{1/2}_{\gamma Z}$ which is generated
from the (tree-level) ${\tilde\chi}_i^0 {\tilde\chi}_1^0Z$ vertex through
a $Z-\gamma$ one-loop transition. This contribution was missing
in~\cite{Ullio:1997ke} (and by extension~\cite{Das:2012ys}) and is gauge-dependent, as was first pointed out
in~\cite{Boudjema:2005hb} for the MSSM calculation of $\svgZ$. Besides being needed to obtain a gauge invariant result, this
contribution can be numerically significant
when the neutralino has a
significant higgsino-fraction, as is the case we study here. This can be understood by the fact that the coupling of the neutralino to the $Z$ boson is proportional to the neutralino's higgsino component.

Our implementation of the cross-section $\svgg$ follows~\cite{Chalons:2011ia}, which relies on
the \texttt{SloopS}
code~\cite{Boudjema:2005hb,Baro:2007em,Baro:2008bg,Baro:2009gn}. The
\texttt{SloopS} code benefits from a non-linear gauge-fixing which enables the
user to check the gauge invariance of the result. Further details about the
numerical implementation for the NMSSM calculation can be found
in~\cite{Chalons:2011ia}. Note that by default no width is implemented in
\texttt{SloopS} since it can spoil gauge invariance. As the width of the singlet $A_S$ is very narrow the
inclusion of the width is completely negligible for the points we consider.
Although the inclusion of the width in principle relevant when very close to the
resonance $2 m_{\nue} = m_A$, these points give a rate that is much too large 
and would already be excluded, as discussed, for instance, in~\cite{Chalons:2011ia}. In addition to the diagram of figure~\ref{fig:schannel} there
are contributions where the charginos running in the loops are
replaced by leptons and quarks. Analytically their contribution is similar to
the formulas in eq.~\eqref{eq:chigamgam} where the term
$\lambda U_{12} V_{12}$ is replaced by the Yukawa
coupling of the corresponding lepton/quark. However, unlike the chargino, these diagrams are suppressed since $m_{l,q} \not\approx m_{\nue}$. These diagrams are therefore sub-dominant. There is also a slight destructive interference between
the resonant type diagrams and box-type ones with charginos inside. However,
these are also numerically sub-dominant.

\section{Analytic expressions for $\langle \sigma v\rangle_{\nue\nue \rightarrow A_S \rightarrow \gamma \gamma}$ and $\langle \sigma v\rangle_{\nue \nue \rightarrow A_S \rightarrow \gamma Z}$}\label{sec:svan}

In this appendix we present the analytic expressions for $\langle \sigma v\rangle_{\gamma \gamma}$ and $\langle \sigma v\rangle_{\gamma Z}$, considering only the contribution from the s-channel psuedoscalar resonance $A_S$, which we assume is a pure singlet state. To derive these results, we adapted the calculations from~\cite{Bergstrom:1997fh} and~\cite{Ullio:1997ke} for $\svgg$ and $\svgZ$ respectively. Unlike our numerical result, our analytic result for $\svgZ$ does not include the field renormalisation $\delta Z_{\gamma Z}^{1/2}$.

Including both charginos in the triangle and assuming that $\lambda$ and $\kappa$ are real, we find that
\begin{equation}
\langle \sigma v\rangle_{\gamma \gamma}=\frac{\alpha^2}{4 \pi^3}\frac{\lambda^2 \left(\lambda N_{13} N_{14}-\kappa N_{15}^2 \right)^2}{(4 m_{\nue}^2-m_A^2)^2+\Gamma_A^2 m_A^2} \,\mathcal{G}_{\tilde{\chi}^{\pm}}^2\;,
\label{eq:fullgg}
\end{equation}
where we have defined
\begin{equation}
\mathcal{G}_{\tilde{\chi}^{\pm}}=\sum_{\tilde{\chi}^+_i} m_{\tilde{\chi}^+_i} U_{i2} V_{i2}\arctan^2{\sqrt{\frac{m_{\nue}^2}{m_{\tilde{\chi}^+_i}-m_{\nue}^2}}}
\end{equation}
and $U$ and $V$ are the usual unitary matrices that diagonalise the chargino mass matrix. 

Similarly, we find
\begin{equation}
\langle \sigma v\rangle_{\gamma Z}=\frac{\alpha^2}{32 \pi^3}\frac{1}{\sin^2 2 \theta_W}\left( 1-\frac{m_Z^2}{4 m_{\nue}^2}\right)\frac{\lambda^2 \left(\lambda N_{13} N_{14}-\kappa N_{15}^2 \right)^2}{(4 m_{\nue}^2-m_A^2)^2+\Gamma_A^2 m_A^2} \,\mathcal{F}_{\tilde{\chi}^{\pm}}^2\;,
\end{equation}
where we have defined
\begin{equation}
\mathcal{F}_{\tilde{\chi}^{\pm}}=\sum_{\tilde{\chi}^+_i,\, \tilde{\chi}^+_j} m_{\tilde{\chi}^+_i} \tilde{S}_{ij} I_1^4\left(\frac{m_{\nue}^2}{m_{\tilde{\chi}^+_i}^2},\frac{m_{\tilde{\chi}^+_j}^2}{m_{\tilde{\chi}^+_i}^2},1,\frac{m_Z^2}{4 m_{\tilde{\chi}^+_i}^2}\right)+m_{\tilde{\chi}^+_j} \tilde{D}_{ij}I_1^4\left(\frac{m_{\nue}^2}{m_{\tilde{\chi}^+_i}^2},1,\frac{m_{\tilde{\chi}^+_j}^2}{m_{\tilde{\chi}^+_i}^2},\frac{m_Z^2}{4 m_{\tilde{\chi}^+_i}^2}\right)\;.
\end{equation}
Here, 
\begin{align}
\tilde{S}_{ij}&=U^{*}_{i2}V^*_{j2}\tilde{O}^{' L}_{ji}+U_{j2}V_{i2}\tilde{O}^{' R}_{ji}\\
\tilde{D}_{ij}&=U^{*}_{i2}V^*_{j2}\tilde{O}^{' R}_{ji}+U_{j2}V_{i2}\tilde{O}^{' L}_{ji}\,
\end{align}
where
\begin{align}
\tilde{O}^{' L}_{ji}&=-V_{j1}V^*_{i1}-\frac{1}{2}V_{j2}V_{i2}^*+\delta_{ij}\sin^2\theta_W\\
\tilde{O}^{' R}_{ji}&=-U_{j1}^*U_{i1}-\frac{1}{2}U^*_{j2}U_{i2}+\delta_{ij}\sin^2\theta_W\;.
\end{align}
Finally, we also have
\begin{equation}
I_1^4(a,b,c,d)=\int_0^1 \frac{dx}{x}\left[s{\rm{log}}(-4a,b,c;x)-s{\rm{log}}(-4d,b,c;x) \right]\;,
\end{equation}
where
\begin{equation}
s{\rm{log}}(a,b,c;x)\equiv\log\left[|-a x^2+(a+b-c)x+c|\right]\;.
\end{equation}

Once $m_{\nue}$ is fixed (and since $m_Z$ and $\sin\theta_W$ are known) the ratio
\begin{equation}
\frac{\svgZ}{\svgg}=\frac{1}{8 \sin^2 2\theta_W}\left( 1-\frac{m_Z^2}{4 m_{\nue}^2}\right) \frac{\mathcal{F}^2_{\tilde{\chi}^{\pm}}}{\mathcal{G}^2_{\tilde{\chi}^{\pm}}}
\end{equation}
depends only on the chargino masses and mixing matrices through $\mathcal{G}_{\tilde{\chi}^{\pm}}$ and $\mathcal{F}_{\tilde{\chi}^{\pm}}$. Therefore, an accurate measurement of this ratio will constrain properties of the chargino sector.


\section{Overview of direct detection results}
\label{sec:ddcalc}

Here we briefly review the computation to calculate the neutralino scattering rate at direct detection experiments. The low energy neutralino-quark effective Lagrangian is
\begin{equation}
\mathcal{L}_{\rm{eff}}= a_q \bar{\tilde{\chi}}_1^0 \nue \bar{q}q+d_q \bar{\tilde{\chi}}_1^0\gamma^{\mu}\gamma_5 \nue \bar{q}\gamma_{\mu}\gamma_5 q\;,
\end{equation}
where we only consider contributions that are not velocity or momentum transfer suppressed. The first-term gives rises to a spin-independent interaction while the second gives rise to a spin-dependent interaction. We consider each in turn.

For spin-independent interactions, experiments typically quote the cross-section to scatter off a nucleon, given by
\begin{equation}
\sigma_{\rm{SI}}=\frac{4 \mu_{N \nue}^2}{\pi}\frac{(Z f_p+(A-Z) f_n)^2}{A^2}\;,
\end{equation}
where $\mu_{N \nue}$ is the neutralino-nucleon reduced mass and $A$ and $Z$ are the nucleon number and charge of the target nucleus. The couplings to protons and neutrons, $f_p$ and $f_n$, are related to the quantities entering $\mathcal{L}_{\rm{eff}}$ by
\begin{equation}
\frac{f_p^N}{m_N}=\sum_{q=u,d,s}\frac{a_q}{m_q}f_{T_q}^N+\frac{2}{27} f_{T_{G}}^N\sum_{Q=c,b,t}\frac{a_Q}{m_Q}\;.
\end{equation}
Here, the $f_{T_q}$ are determined from lattice QCD and chiral perturbation theory and $f_{T_G}=1-\sum_{q=u,d,s}f_{T_q}$~\cite{Shifman:1978zn}. Unless stated otherwise, we take $\sigma_0=35$~MeV and $\sigma_{\pi N}=45$~MeV, leading to
\begin{align}
& f_{T_u}^p=0.020, \qquad f_{T_d}^p=0.026,\qquad f_{T_s}^p=0.13 \\
& f_{T_u}^n=0.014, \qquad f_{T_d}^n=0.036, \qquad f_{T_s}^n=0.13 
\end{align}
These values are comparable to the values used in DarkSUSY~\cite{Gondolo:2004sc}. The uncertainty associated with the strange quark matrix element $f_{T_s}$ is large; more recent values are typically much lower.  For instance, ref.~\cite{Cheng:2012qr} find $f_{T_s}^p=0.053$ which leads to a cross-section that is smaller by a factor of a few (comparable values of $f_{T_s}$ were also found in~\cite{Takeda:2010cw, Alarcon:2011zs, Thomas:2012tg, Alarcon:2012nr}). 

In principle, $a_q$ receives contributions from the t-channel exchange of CP-even Higgs bosons and s-channel squark exchange. However, for the discussion below, we ignore the contribution from the squarks because of the strong LHC limits on first two generations of squarks~\cite{ATLAS-CONF-2012-109,:2012mfa} (in the absence of unusual features in the spectrum such as compression or $R$-parity violation which are not present in our model). In this case, for u-type quarks,
\begin{equation}
\frac{a_u}{m_u}=\frac{g}{4 m_W \sin\beta }\sum_{j=1}^3 \frac{g_{h_j \chi \chi} S_{j2}}{m_{h_j}^2}\;.
\end{equation}
The corresponding expression for $a_d/m_d$ can be found by replacing $\sin \beta \rightarrow \cos\beta$ and $S_{j2}\rightarrow S_{j1}$. The matrix $S_{ij}$ diagonalises the CP Higgs mass matrix such that the mass eigenstates $h_i$, ordered such that $m_{h_1}<m_{h_2}<m_{h_3}$, are related to the weak eigenstates $h_i^{\rm{weak}}=(H_{dR},H_{uR}, S_R)$, where $\sqrt{2} H_{a}=H_{aR}+iH_{aI}$, through $h_i=S_{ij} h_j^{\rm{weak}}$. The coupling $g_{h_i \chi \chi}$ for the lightest neutralino is
\begin{align}
\begin{split}
g_{h_i \chi \chi}=&g(N_{12}-\tan\theta_W N_{11})(S_{i1}N_{13}-S_{i2}N_{14})\\
&-\sqrt{2}\lambda(S_{i1}N_{14}N_{15}+S_{i2}N_{13}N_{15}+S_{i3}N_{13}N_{14})\\
&+\sqrt{2}\kappa S_{i3} N_{15}^2\;.
\end{split}
\end{align}
The first line reduces to the MSSM contribution while the second and third lines come from the $\lambda S H_u H_d$ and $\frac{\kappa}{3}S^2$ superpotential terms respectively.

We next consider the spin-dependent interaction. The additional terms do not give rise to any new terms not suppressed by velocity or the momentum-transfer so the result is the same as for the MSSM. The dominant contribution comes from t-channel $Z$-exchange so couples directly the higgsino component. The cross-section is
\begin{equation}
\sigma_{p,n}^{\rm{SD}}=\frac{12 \mu_{p,n \nue}^2}{\pi} \left(\sum_{q=u,d,s} d_q \Delta_q^{p,n} \right)^2 \;,
\end{equation}
where
\begin{equation}
d_q=\frac{g^2}{4 m_W^2}\frac{T_{3q}}{2}\left( |N_{13}|^2-|N_{14}|^2\right)
\end{equation}
and $T_{3q}$ is the third component of hypercharge. The fraction of the nucleon spin carried by a given quark is given by $\Delta_q^{p,n}$ and we use
\begin{equation}
\Delta_u^p=\Delta_d^n=0.842,\qquad \Delta_d^p=\Delta_u^n=-0.427, \qquad \Delta_s^p=\Delta_s^n=-0.085\,.
\end{equation}

With these numbers, we find that the cross-section to scatter off a proton is larger than the cross-section to scatter off a neutron:
\begin{align}
\sigma_{p}^{\rm{SD}}&\approx 4.0\times10^{-4}\text{ pb}\left( \frac{|N_{13}|^2-|N_{14}|^2}{0.1}\right)^2 \\
\sigma_{n}^{\rm{SD}}&\approx 3.1\times10^{-4}\text{ pb}\left( \frac{|N_{13}|^2-|N_{14}|^2}{0.1}\right)^2\;.
\end{align}

\bibliography{ref}
\bibliographystyle{ArXiv}

\end{document}